\documentclass[twocolumn]{jpsj2}
\def\Ms{M_{\rm s}}
\def\Nc{{N_{\rm c}}}

\def\H{{\cal H}}
\def\r{{\rm R}}
\def\l{{\rm L}}
\def\d{{\rm d}}
\def\e{{\rm e}}
\def\o{{\rm o}}
\def\v#1{\mib #1}
\def\img{{\rm i}} %
\def\vs#1{\mbox{\scriptsize\boldmath$#1$}}
\newlength{\figurewidth}
\def\runauthor{Kazuo {\sc Hida}, Masaru {\sc Shiino} and Wei {\sc Chen}}

\title
{
Bond Operator Mean Field Approach to the Magnetization Plateaux in Quantum Antiferromagnets \\ {\it - Application to the  S=1/2 Coupled Dimerized Zigzag Heisenberg Chains - }
}

\author
{Kazuo {\sc Hida}\thanks{e-mail: hida@phy.saitama-u.ac.jp}, Masaru {\sc Shiino}\thanks{Present address: Toyota Techno Service Corp., 2-88, Hoei-cho, Toyota, Aichi, 470-1201} and Wei {\sc Chen}$^{1, }$\thanks{Present address: Institute of Biomedical Engineering, University of Montreal, Montreal, Quebec, Canada H4J 1C5}}

\inst
{
Department of Physics, Faculty of Science, \\ Saitama University, Saitama 338-8570 \\
$^1$ Department of Chemistry, McGill University, Montreal, PQ, Canada H3A 2K6 
}

\recdate
{
January 31, 2004
}

\abst
{
The magnetization plateaux in two dimensionally coupled $S=1/2$ dimerized zigzag Heisenberg chains are investigated by means of the bond operator mean field approximation. In the absence of the interchain coupling, this model is known to have a plateau at half of the saturation magnetization accompanied by the spontaneous translational symmetry breakdown. The parameter regime in which the plateau appears is reproduced well within the present approximation. In the presence of the interchain coupling, this plateau is shown to be suppressed. This result is also supported by the numerical diagonalization calculation.
}
\kword
{
zigzag Heisenberg chain, dimerization, magnetization plateau, bond operator mean field approximation
}

\begin{document}
\maketitle
\section{Introduction}
The phenomenon of the magnetization plateau in quantum magnets has been  attracting broad interest as an essentially macroscopic quantum phenomenon in which macroscopic magnetization $M$ is quantized to the fractional values of the saturated magnetization $\Ms$.\cite{mo, hk, tone, tone1, totsuka, kole, matsu, oku, oku2, naru, WS, ketal, hoso} Among them, the plateau associated with the spontaneous translational symmetry breakdown (STSB) is most interesting because such a plateau is the manifestation of the  many body effect.\cite{totsuka}

 Theoretically, the plateaux in one dimensional quantum magnets are best understood, because the size problem inherent to numerical studies are less serious and the analytical tools such as bosonization and conformal field theory are also available. On the other hand, in real materials, the interchain coupling is inevitable.\cite{naru, WS, ketal, hoso} However, the limitation of the system size  becomes serious in the numerical studies of the systems in higher dimensions. Therefore it is desirable to to develop a reliable approximation scheme appropriate for the description of the plateau state in higher dimensions. 

Among various approximation schemes which are commonly used in quantum spin systems, the bond operator mean field  approximation (BOMFA) is known to be powerful for the description of the spin gap state in the {\it absence} of magnetization.\cite{sb, grs, mgc, ygws} Considering that the plateau state can be  regarded as a spin gap state with non-zero magnetization, this method must be also useful  for the description of the plateau state. Actually, there are several attempts to describe the magnetized states in terms of the bond operators.\cite{matsu,svb,mnrs,wls,ymmi} In the present work, we propose the application of the BOMFA to the plateau state in the two dimensionally coupled $S=1/2$ dimerized zigzag Heisenberg  chains. Especially, we show that the plateau accompanied by the STSB can be described by the BOMFA with corresponding configuration of singlet and triplet pairs.

The single $S=1/2$ dimerized zigzag Heisenberg chain is known to have a magnetization plateau at $m (\equiv M/\Ms)=1/2$ in the appropriate parameter range.\cite{tone, tone1, totsuka} Our approximation {\it quantitatively} reproduce the phase diagram at $m=1/2$. Motivated by the success in one dimension, we further investigate the effect of the interchain coupling within our approximation scheme. It is found that the interchain coupling suppresses this plateau. The numerical diagonalization results also supports this conclusion.  

This paper is organized as follows. In the next section, we briefly review the bond operator transformation. The Hamiltonian of the two dimensionally coupled $S=1/2$ dimerized zigzag Heisenberg  chains is presented and rewritten using the bond operators in \S 3. In section 4, the mean field approximation appropriate for the plateau state is introduced and applied for the present Hamiltonian. The condition for the stability of the plateau is derived. The ground state phase diagram is presented in \S 5. The last section is devoted to summary and discussion.

\section{Bond Operators}
The bond operators  $s$, $s^{\dagger }$, $t_{\alpha}$  and $t^{\dagger }_{\alpha} (\alpha=\pm, 0)$ are defined for a pair of two $S=1/2$ spins $\v{S}_\l$ and  $\v{S}_\r$ as\cite{sb, grs, mgc, ygws}
\begin{eqnarray}
\label{op1}s^{\dagger }\left.|0\right\rangle&=&\left.|s\right\rangle=\frac{1}{\sqrt{2}}\left(\left.|\uparrow \downarrow \right\rangle-\left.|\downarrow \uparrow \right\rangle\right), \\
\label{op2}t_{+}^{\dagger }\left.|0\right\rangle&=&\left.|t_{+}\right\rangle=-\left.|\uparrow \uparrow \right\rangle, \\
\label{op3}t_{-}^{\dagger }\left.|0\right\rangle&=&\left.|t_{-}\right\rangle=\left.|\downarrow \downarrow \right\rangle, \\
\label{op4}t_{0}^{\dagger }\left.|0\right\rangle&=&\left.|t_{0}\right\rangle=\frac{1}{\sqrt{2}}\left(\left.|\uparrow \downarrow \right\rangle+\left.|\downarrow \uparrow \right\rangle\right), 
\end{eqnarray}
where  $\left.|0\right\rangle$ is the vacuum  state and  $\left.|S_\l^z S_\r^z \right\rangle$ is the eigenstate of $S_\l^z$ and  $S_\r^z$.  These operators  satisfy the bosonic commutation relations, 
\begin{equation}
\left[s, s^{\dagger }\right]=1, \, \left[t_{\alpha}, t_{\beta}^{\dagger }\right]=\delta _{\alpha\beta }, \, \left[s, t_{\alpha}^{\dagger }\right]=0.
\end{equation}
Because the physical state of each spin pair is either  singlet  or one of the three triplet states, these boson operators  must satisfy the following condition, 
\begin{equation}
\label{con}s^{\dagger }s+t_{+}^{\dagger }t_{+}+t_{-}^{\dagger }t_{-}+t_{0}^{\dagger }t_{0}=1.
\end{equation}
The components of the spin operators ${\v{S}}_{\r}$ and ${\v{S}}_{\l}$ are rewritten in terms of the bond operators as follows:
\begin{eqnarray}
&&S_{\r, \l}^+=\frac{1}{\sqrt{2}}\left\{\pm\left(s^{\dagger }t_{-}+{t_{+}}^{\dagger }s\right)-\left(t_{+ }^{\dagger }t_{0 }-t_{0}^{\dagger }t_{-}\right)\right\}, \hspace{5mm} \\
&&S_{\r, \l}^-=\frac{1}{\sqrt{2}}\left\{\pm\left(s^{\dagger }t_{+}+{t_{-}}^{\dagger }s\right)+\left(t_{- }^{\dagger }t_{0 }-t_{0}^{\dagger }t_{+}\right)\right\}, \\
&&S_{\r, \l}^z=\frac{1}{2}\left\{\pm\left(s^{\dagger }t_{0}+{t_{0}}^{\dagger }s\right)+\left(t_{+ }^{\dagger }t_{+ }-t_{-}^{\dagger }t_{-}\right)\right\}, 
\end{eqnarray}
where upper and lower sign in rhs correspond to $\r$ and $\l$ in lhs, respectively.

\section{Hamiltonian}
\begin{figure}
  \centerline{\includegraphics[width=\figurewidth]{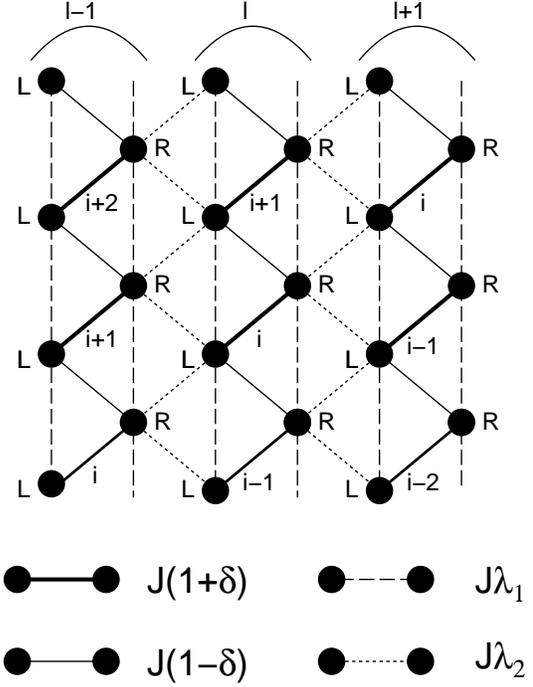}}
  \caption{Two dimensionally coupled dimerized zigzag chains. The filled circles represent the $S=1/2$ spins.}%
\label{lattice}
  \end{figure}

The Hamiltonian of the two dimensionally coupled $S=1/2$ dimerized zigzag chains in the magnetic field $H$ along $z$-direction  is given by, 

\begin{eqnarray}
\lefteqn{\H=J\sum _{l=1}^{\Nc}\sum _{i=1}^{N}\left\{\left(1+\delta\right){\v{S}}_{\l}(i, l)\cdot {\v{S}}_{\r}(i, l)\right.}\nonumber\\
&+&\left(1-\delta\right){\v{S}}_{\l}(i, l)\cdot {\v{S}}_{\r}(i+1, l)  \nonumber \\
&+&\lambda _{1}\left({\v{S}}_{\l}(i, l)\cdot {\v{S}}_{\l}(i+1, l)+{\v{S}}_{\r}(i, l)\cdot {\v{S}}_{\r}(i+1, l)\right)\nonumber\\
&+&\lambda _{2}\left({\v{S}}_{\r}(i, l)\cdot {\v{S}}_{\l}(i, l+1)\right.\nonumber\\
&+&\left.\left.{\v{S}}_{\r}(i, l)\cdot {\v{S}}_{\l}(i+1, l+1)\right)\right\}\nonumber\\
&-&\sum _{l=1}^{\Nc}\sum _{i=1}^{N}H\left(S_{\l}^z(i, l)+ S_{\r}^z(i, l)\right), 
\label{ham}
\end{eqnarray}
where $N$ is the length of each chain and $\Nc$ is the number of chains.  The chains are distinguished by the index $l$. The dimers coupled via the stronger nearest neighbour bonds $J(1+\delta)$ are the elementary components of the present model. In the following, these spin pairs are simply called 'dimers'. They are  distinguished by the index $i$ in each chain. The indices $\r$ and $\l$ distinguish the two spins in each dimer. Accordingly, the spin operators ${\v{S}}_{\r}(i, l)$ and ${\v{S}}_{\l}(i, l)$ are the right and left spins in the $i$-th dimer on the $l$-th chain, respectively. The weaker intrachain nearest neighbour coupling, the intrachain next nearest neighbour coupling and  the interchain coupling are denoted by $J(1-\delta)$, $J\lambda_1$ and  $J\lambda_2$, respectively, as depicted in Fig. \ref{lattice}.  This model can be also regarded as a  dimerized triangular Heisenberg model. The uniform triangular lattice corresponds to $\lambda_1=\lambda_2=1$ and $\delta=0$.

The Hamiltonian (\ref{ham}) can be represented by the bond operators for the dimers connected by $J(1+\delta)$ bonds as follows, 
\begin{equation}
\label{2dham} \H=\sum _{j=0}^{4}\H_{j}, 
\end{equation}
\begin{eqnarray}
\lefteqn{\H_{0}=\sum _{l=1}^{\Nc}\sum _{i=1}^{N}\left\{J\left(1+\delta \right)\left(-\frac{3}{4}s^{\dagger}(i, l)s(i, l)\right.\right.}\nonumber\\
&+&\left.\frac{1}{4}\sum _{\alpha=0, \pm}t_{\alpha}^{\dagger}(i, l)t_{\alpha}(i, l)\right)-\mu(i, l)\Big( s^{\dagger}(i, l)s(i, l)\nonumber\\
&+&\left.\sum _{\alpha=0, \pm}t_{\alpha}^{\dagger}(i, l)t_{\alpha}(i, l)-1\Big)\right\}, \\
\lefteqn{\H_{1}=-\sum _{l=1}^{\Nc}\sum _{i=1}^{N}\frac{J}{4}\left(1-\delta\right)\left[\left(s^{\dagger}(i, l)t_{0}(i, l)+t_{0}^{\dagger}(i, l)s(i, l)\right.\right.}\nonumber\\
&+&\left.t_{+}^{\dagger}(i, l)t_{+}(i, l)-t_{-}^{\dagger}(i, l)t_{-}(i, l)\right)\nonumber\\
&\times&\left(s^{\dagger}(i+1, l)t_{0}(i+1, l)+t_{0}^{\dagger}(i+1, l)s(i+1, l)\right.\nonumber\\
&-&\left.t_{+}^{\dagger}(i+1, l)t_{+}(i+1, l)+t_{-}^{\dagger}(i+1, l)t_{-}(i+1, l)\right)\nonumber\\
&+&\left\{\left(s^{\dagger}(i, l)t_{-}(i, l)+t_{+}^{\dagger}(i, l)s(i, l)\right.\right.\nonumber\\
&+&\left.t_{0}^{\dagger}(i, l)t_{-}(i, l)-t_{+}^{\dagger}(i, l)t_{0}(i, l)\right)\nonumber\\
&\times&\left(s^{\dagger}(i+1, l)t_{+}(i+1, l)+t_{-}^{\dagger}(i+1, l)s(i+1, l)\right.\nonumber\\
&-&\left.\left.\left.t_{-}^{\dagger}(i+1, l)t_{0}(i+1, l)+t_{0}^{\dagger}(i+1, l)t_{+}(i+1, l)\right) \right.\right.\nonumber\\
&+&{\rm h.c.}\Big\}\Big], \\
\lefteqn{\H_{2}=\sum _{l=1}^{\Nc}\sum _{i=1}^{N}\frac{J}{2}\lambda_1\left[\left(s^{\dagger}(i, l)t_{0}(i, l)+t_{0}^{\dagger}(i, l)s(i, l)\right)\right.}\nonumber\\
&\times&\left(s^{\dagger}(i+1, l)t_{0}(i+1, l)+t_{0}^{\dagger}(i+1, l)s(i+1, l)\right)\nonumber\\
&+&\left(t_{+}^{\dagger}(i, l)t_{+}(i, l)-t_{-}^{\dagger}(i, l)t_{-}(i, l)\right)\nonumber\\
&\times&\left(t_{+}^{\dagger}(i+1, l)t_{+}(i+1, l)-t_{-}^{\dagger}(i+1, l)t_{-}(i+1, l)\right)\nonumber\\
&+&\left\{\left(s^{\dagger}(i, l)t_{-}(i, l)+t_{+}^{\dagger}(i, l)s(i, l)\right)\right.\nonumber\\
&\times&\left(s^{\dagger}(i+1, l)t_{+}(i+1, l)+t_{-}^{\dagger}(i+1, l)s(i+1, l)\right)\nonumber\\
&+&\left(t_{0}^{\dagger}(i, l)t_{-}(i, l)-t_{+}^{\dagger}(i, l)t_{0}(i, l)\right)\nonumber\\
&\times&\left.\left.\left(t_{0}^{\dagger}(i+1, l)t_{+}(i+1, l)-t_{-}^{\dagger}(i+1, l)t_{0}(i+1, l)\right)\right.\right.\nonumber\\
&+&{\rm h.c.}\Big\}\Big], \\
\lefteqn{\H_{3}=-\sum _{l=1}^{\Nc}\sum _{i=1}^{N}\frac{J}{4}\lambda_2\left[\left(s^{\dagger}(i, l)t_{0}(i, l)+t_{0}^{\dagger}(i, l)s(i, l)\right.\right.}\nonumber\\
&+&t_{+}^{\dagger}(i, l)t_{+}(i, l)-t_{-}^{\dagger}(i, l)t_{-}(i, l)\nonumber\\
&+&s^{\dagger}(i+1, l)t_{0}(i+1, l)+t_{0}^{\dagger}(i+1, l)s(i+1, l)\nonumber\\
&+&t_{+}^{\dagger}(i+1, l)t_{+}(i+1, l)-\left.t_{-}^{\dagger}(i+1, l)t_{-}(i+1, l)\right)\nonumber\\
&\times&\left(s^{\dagger}(i, l+1)t_0(i, l+1)+t_0^{\dagger}(i, l+1)s(i, l+1)\right.\nonumber\\
&-&\left.t_+^{\dagger}(i, l+1)t_+(i, l+1)+t_-^{\dagger}(i, l+1)t_-(i, l+1)\right)\nonumber\\
&+&\left\{\left(s^{\dagger}(i, l)t_{-}(i, l)+t_{+}^{\dagger}(i, l)s(i, l)\right.\right.\nonumber\\
&-&\left.\left.t_{0}^{\dagger}(i, l)t_{-}(i, l)+t_{+}^{\dagger}(i, l)t_{0}(i, l)\right.\right.\nonumber\\
&+&s^{\dagger}(i+1, l)t_{-}(i+1, l)+t_{+}^{\dagger}(i+1, l)s(i+1, l)\nonumber\\
&-&\left.t_{0}^{\dagger}(i+1, l)t_{-}(i+1, l)+t_{+}^{\dagger}(i+1, l)t_{0}(i+1, l)\right)\nonumber\\
&\times&\left(s^{\dagger}(i, l+1)t_+(i, l+1)+t_-^{\dagger}(i, l+1)s(i, l+1)\right.\nonumber\\
&-&\left.\left.\left.t_0^{\dagger}(i, l+1)t_+(i, l+1)+t_-^{\dagger}(i, l+1)t_0(i, l+1)\right)\right.\right.\nonumber\\
&+&{\rm h.c.}\Big\}\Big], \\
\lefteqn{\H_{4}=-\sum _{l=1}^{\Nc}\sum _{i=1}^{N}H\left(t_{+}^{\dagger}(i, l)t_{+}(i, l)-t_{-}^{\dagger}(i, l)t_{-}(i, l)\right), }
\end{eqnarray}
where the Lagrange multipliers $\mu(i, l)$ are introduced to account for the constraint (\ref{con}).

\section{Bond Operator Mean Field Approximation in the Plateau State}

\begin{figure}
  \centerline{\includegraphics[width=0.8\figurewidth]{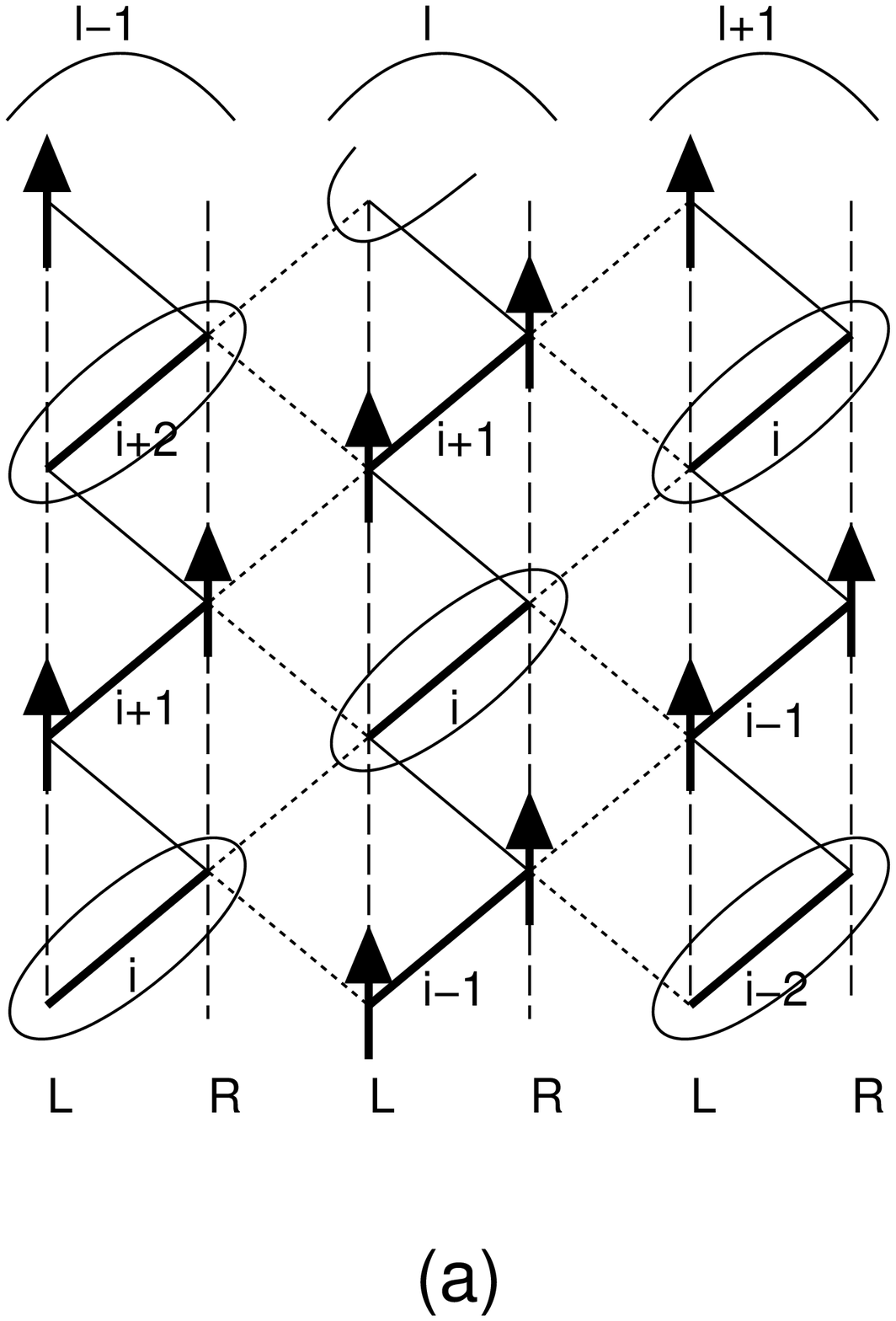}}
  \centerline{\includegraphics[width=0.8\figurewidth]{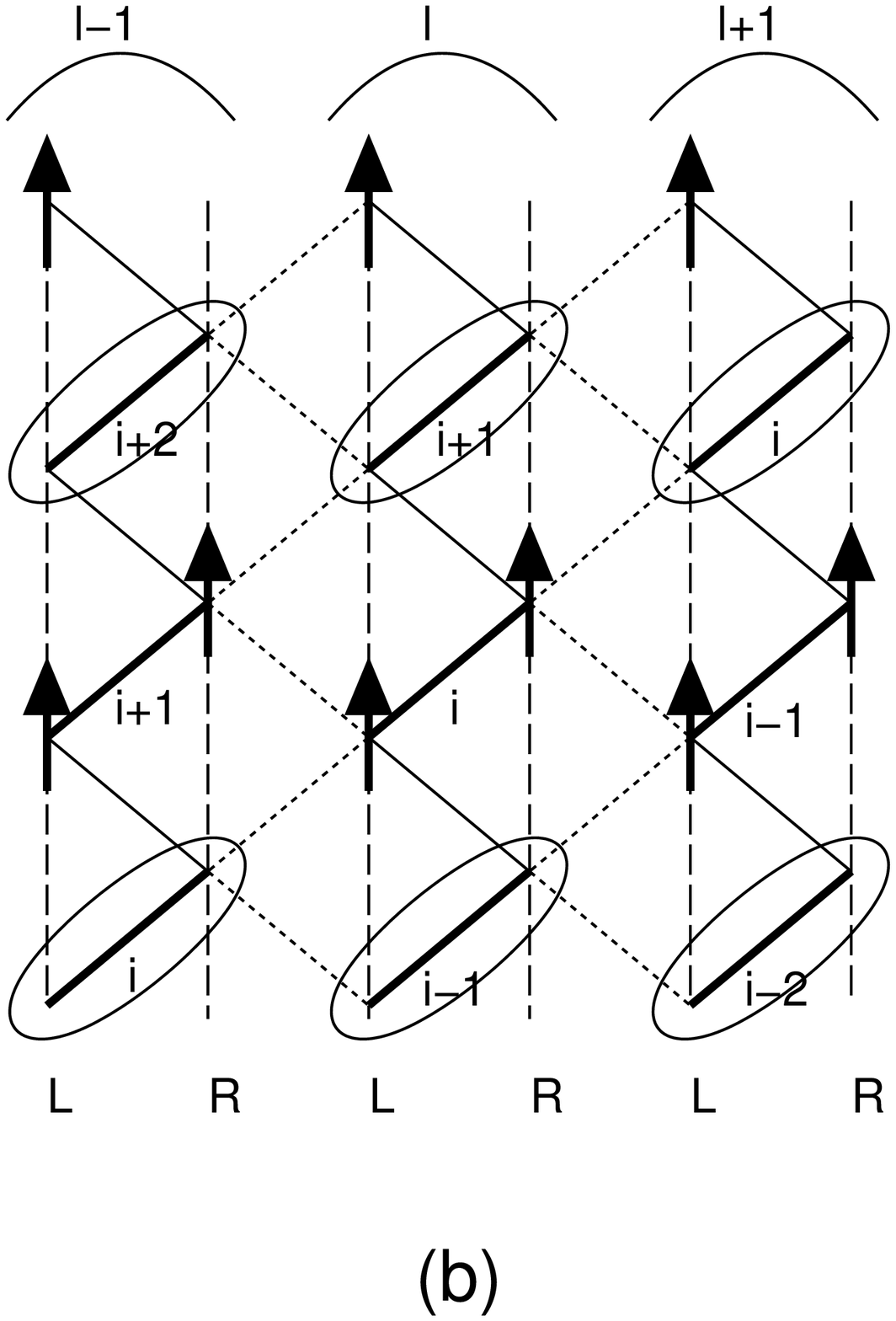}}
  \caption{Two possible configurations of the triplet and singlet pairs in the $m=1/2$ plateau state. The spin pairs in each oval form singlet pairs and the spins represented by the thick up arrows are spins in the triplet state polarized along the magnetic field. }%
\label{conf}
  \end{figure}

On the $m=1/2$ plateau state, it is expected that half of the dimers on the $J(1+\delta)$-bonds are in the singlet states and others are in the triplet states polarized along the $z$-direction as far as $1+\delta >> 1-\delta, \lambda_1, \lambda_2$. In this state, the translational symmetry $i \rightarrow i+1$ is spontaneously broken to $i \rightarrow i+2$. In the two dimensional case, there are two possible singlet-triplet configurations as depicted in Fig. \ref{conf}(a) and (b). We have carried out our calculation for both configurations. However, it turned out that the configuration (a) gives wider plateau. Therefore, we only present the calculation for the configuration  (a) in the following. The calculation for the configuration (b) can be carried out almost in the same manner. 

Based on the above explained physical picture of the plateau state, we assume the condensation of $s(2i, l)$ and $t_+(2i+1, l)$ bosons and keep the ground state expectation values of these operators as $s=<s^{\dagger }(2i, l)>=<s(2i, l)>$ and  $t=<t^{\dagger}_{+}(2i+1, l)>=<t_{+}(2i+1, l)>$. 

We also make global approximation for the Lagrange multiplier $\mu(i, l)$. In the present case, the $(2i, l)$-th sites and the $(2i+1, l)$-th sites are inequivalent. Therefore we assume different values $\mu^{\e}$ and $\mu^{\o}$ for $\mu(2i, l)$ and $\mu(2i+1, l)$, respectively. Thus, the ground state energy is obtained as, 
\begin{eqnarray}
\lefteqn{E_0=\frac{\Nc N}{2}\left\{J\left(-\frac{3}{4}s^2+\frac{1}{4}t^2\right)(1+\delta)  \right.}\nonumber\\
&+& \left.\frac{J\lambda_2}{4}t^4-Ht^2+\mu^{\o}(t^2-1)+\mu^{\e}(s^2-1)\right\}
\end{eqnarray}
neglecting all quantum fluctuation terms. 

The parameters $s$, $t$, $\mu^{\e}$ and $\mu^{\o}$ are determined so as to optimize $E_0$ as follows, 
\begin{eqnarray}
\label{mu1}\mu^{\o}&=&-\frac{J}{4}\left(1+\delta\right)-\frac{J\lambda_2}{2}+H, \\
\label{mu2}\mu^{\e}&=&\frac{3}{4}J\left(1+\delta\right), \\
s&=&t=1.
\end{eqnarray}

Neglecting the terms higher than the third order in ${t}_{\pm}(2i, l)$, ${t}_{0}(2i, l)$, ${s}(2i+1, l)$, ${t}_{0}(2i+1, l)$ and ${t}_-(2i+1, l)$ and carrying out the Fourier transformation, 
\begin{eqnarray}
&&t_{\alpha}(2i, l)=\sqrt{\frac{2}{\Nc N}}\sum _{{\vs{k}}}t^{\e}_{\alpha}(\v{k}) {\e}^{{\img} {\vs{k}}\cdot {\vs{r}(2i, l)}}, \ (\alpha=0, \pm),\nonumber\\
&& \\
&&{s}(2i+1, m)=\sqrt{\frac{2}{\Nc N}}\sum _{{\vs{k}}}s^{\o}(\v{k}) {\e}^{{\img} {\vs{k}}\cdot {\vs{r}(2i+1, l)}}, \\
&&t_{\alpha}(2i+1, l)=\sqrt{\frac{2}{\Nc N}}\sum _{{\vs{k}}}t^{\o}_{\alpha}(\v{k}) {\e}^{{\img} {\vs{k}}\cdot {\vs{r}(2i+1, l)}}, \nonumber\\
&&\hspace{5cm} (\alpha=0, -), 
\end{eqnarray}
where $\v{r}(i, l)$ is the position of the $(i, l)$-th dimer, we obtain the BOMFA Hamiltonian $\H^{\rm MF}$ as, 
\begin{eqnarray}\nonumber
\lefteqn{\H^{\rm MF}=E_{0}}\nonumber\\
&+&\sum _{{\vs{k}}}\left(A_{+}^{\e}(\v{k})t_{+}^{\e\dagger }(\v{k})t^{\e}_{+}(\v{k})\right.
+A_{-}^{\e}(\v{k}) t_{-}^{\e\dagger }(\v{k})t_{-}^{\e}(\v{k})  \nonumber \\
&+&\left.A_{0}^{\e}(\v{k}) t_0^{\e\dagger }(\v{k})t_{0}^{\e}(\v{k})+A_{s}^{\o}(\v{k})s_{+} ^{\o\dagger }(\v{k})s^{\o}_{+}(\v{k})\right.\nonumber\\
&+&\left.A_{-}^{\o}(\v{k}) t^{\o\dagger }_{-}(\v{k})t_{-}^{\o}(\v{k})+A_{0}^{\o}(\v{k}) t_{0} ^{\o\dagger }(\v{k})t_{0}^{\o}(\v{k})\right)  \nonumber\\
&+ &\sum _{{\vs{k}}}P(\v{k})\left\{s^{\o\dagger }(\v{k}) \left(t_{+}^{\e\dagger}(-\v{k})+t_{-}^{\e}(\v{k})\right)\right.\nonumber\\
&+&\left.s^{\o}(\v{k})\left(t_+^{\e}(-\v{k})+t_{-}^{\e\dagger}(\v{k})\right)\right\}  \nonumber \\ 
&+&\sum _{{\vs{k}}}iQ(\v{k})\left\{t_{0}^{\o\dagger }(\v{k})\left(t_{+}^{\e\dagger}(-\v{k})+t_{-}^{\e}(\v{k})\right)\right.\nonumber\\
&-&\left.t_{0}^{\o}(\v{k})\left(t_{+}^{\e}(-\v{k})+t_{-}^{\e\dagger}(\v{k})\right)\right\}\nonumber  \\ 
&+&\sum _{{\vs{k}}}R(\v{k})\Big\{t_{-}^{\e}(\v{k})t_{+}^{\e}(-\v{k})+t_{-}^{\e\dagger}(\v{k})t_{+}^{\e\dagger}(-\v{k})\nonumber\\
&+&\left.\frac{1}{2}\left(t_0^{\e}(\v{k})t_{0}^{\e}(-\v{k})+t_{0} ^{\e\dagger}(\v{k})t_{0}^{\e\dagger}(-\v{k})\right)\right\} \nonumber\\
&+&\sum _{{\vs{k}}}iS(\v{k})\left(s^{\o\dagger }(\v{k})t_{0} ^{\o}(-\v{k})-t_{0}^{\o\dagger}(-\v{k})s ^{\o}(\v{k})\right), 
\label{1dham}
\end{eqnarray}
where
\begin{eqnarray*}
E_0&=&\frac{\Nc N}{2}\left\{-\frac{J}{2}(1+\delta) + \frac{J\lambda_2}{4}-H\right\}, \\
A_{+}^{\e}(\v{k})&=&J\left(1+\delta \right)-H+J\lambda_1+\frac{J}{2}(1-\delta)\nonumber\\
&+&\frac{J\lambda_2}{2}(1-\cos k_y), \\
A_{-}^{\e}(\v{k})&=&J\left(1+\delta \right)+H-J\lambda_1-\frac{J}{2}(1-\delta)\nonumber\\
&-&\frac{J\lambda_2}{2}(1+\cos k_y), \\
A_{0}^{\e}(\v{k})&=&J\left(1+\delta\right)-\frac{J\lambda_2}{2}\cos k_y, \\
A_{s}^{\o}(\v{k})&=&-J\left(1+\delta\right)+H-\frac{J\lambda_2}{2}(1+\cos k_y), \\
A_{0}^{\o}(\v{k})&=&H-\frac{J\lambda_2}{2}(1-\cos k_y), \\
A_{-}^{\o}(\v{k})&=&2H-J\lambda_2, \\
P(\v{k})&=&-\left(\frac{1-\delta}{2}-\lambda_1\right) J\cos k_x -\frac{J\lambda_2}{2}\cos (k_x-k_y), \\
Q(\v{k})&=&\frac{1-\delta}{2} J\sin k_x -\frac{J\lambda_2}{2}\sin (k_x-k_y), \nonumber\\
S(\v{k})&=&\frac{J\lambda_2}{2}\sin k_y, \ \ R(\v{k})=-\frac{J\lambda_2}{2}\cos k_y, 
\end{eqnarray*}
where $k_x$ and $k_y$ are the momentum components conjugate to $i$ and $l$.

In the matrix representation, eq. (\ref{1dham}) is rewritten as, 
\begin{eqnarray}
\label{gyouham1}\H^{\rm MF}&=&E_{0}+{\sum_{\vs{k}}}'{\v{u}}^{\dagger }(\v{k})D(\v{k}){\v{u}}(\v{k})\nonumber\\
&+&\frac{1}{2}\sum _{{\vs{k}}}\left(A_{+}^{\e}(\v{k})+A_{-}^{\e}(\v{k}) +A_{0}^{\e}(\v{k}) \right.\nonumber\\
&+&\left.A_{s}^{\o}(\v{k})+A_{-}^{\o}(\v{k})+A_{0}^{\o}(\v{k})\right), 
\end{eqnarray}
where
\begin{displaymath}
\v{u}^{\dagger }(\v{k})=\left(  \begin{array}{cc}
    \v{t}^{\dagger}(\v{k})   &  \v{t}(-\v{k})   
  \end{array}
\right)
\end{displaymath}
with 
\begin{eqnarray}
\lefteqn{\v{t}^{\dagger }(\v{k})=}\nonumber\\
&&\hspace{-1cm}\left(
  \begin{array}{cccccc}
    t_{+}^{\e\dagger }(\v{k}) &  t_0^{\e\dagger }(\v{k})  & t_{-}^{\e\dagger }(\v{k})   & s ^{\o\dagger }(\v{k})   & t_{0} ^{\o\dagger }(\v{k})   & t_{-} ^{\o\dagger}(\v{k})  
  \end{array}
\right), \nonumber\\
\end{eqnarray}
and
\begin{eqnarray}
\label{gyouretu1} \lefteqn{D^{\d}(\v{k})=}\nonumber\\
&&\hspace{-1cm}\left(
  \begin{array}{cccccc}
    A_{+}^{\e}(\v{k})   &  0  & 0   &  0  & 0   & 0   \\
     0  & A_{0}^{\e}(\v{k})  & 0   &  0  & 0   & 0    \\
     0  & 0   & A_{-}^{\e}(\v{k})   & P(\v{k})   &  -\img Q(\v{k})   & 0    \\
     0  & 0   & P(\v{k})   &  A_{s}^{\o}(\v{k})  & -\img S(\v{k})   & 0      \\
     0  & 0   & \img Q(\v{k})   & \img S(\v{k})   & A_{0}^{\o}(\v{k})   & 0       \\
     0  & 0   & 0   & 0   & 0   & A_{-}^{\o}(\v{k})   
  \end{array}
\right), \nonumber\\
\end{eqnarray}
\begin{eqnarray}
\label{gyouretu2} D^{\o}(\v{k})=\left(
\begin{array}{cccccc}
     0  &  0  & R(\v{k})   & P(\v{k})   & -\img Q(\v{k})   & 0   \\
     0  & R(\v{k})   & 0   &  0  & 0   & 0    \\
     R(\v{k})  & 0   & 0   &  0  & 0   & 0    \\
     P(\v{k})  & 0   & 0   &  0  & 0   & 0      \\
     \img Q(\v{k}) & 0   & 0   & 0   & 0  & 0       \\
     0  & 0   & 0   & 0   & 0   & 0   
  \end{array}
\right),\hspace{-1.5cm}\nonumber\\
\end{eqnarray}
\begin{equation}
\label{gyouretu} D(\v{k})=\left(
  \begin{array}{cc}
    D^{\rm d}(\v{k})  &  D^{\o}(\v{k})     \\
     D^{\o}(-\v{k})  & D^{\rm d}(-\v{k})
  \end{array}
\right).
\end{equation}
The summation $\displaystyle {\sum_{\vs{k}}}'$ is taken over a half of the Brillouin zone (say $k_x >0$). It is known that this type of Hamiltonian is diagonalizable by the Bogoliubov transformation as, 
\begin{equation}
{\sum _{{\vs{k}}}}'{\v{t}}^{\dagger }(\v{k})D(\v{k}){\v{t}}(\v{k})=\sum _{{\vs{k}}}\sum _{\mu=1, 6}\hbar \omega_{\mu}(\v{k})\left({\alpha}^{\dagger}_{\mu}(\v{k}){\alpha}_{\mu}(\v{k})+\frac{1}{2}\right)
\end{equation}
with six modes $\alpha_{\mu}(\v{k}), (\mu=1, .., 6)$ which have positive excitation energies $\hbar \omega_{\mu}(\v{k})$ if and only if the matrix $D(\v{k})$ is positive definite for all allowed values of $\v{k}$.\cite{colpa} From this condition, the plateau region is determined numerically.

\section{Results}
\subsection{Single Dimerized Zigzag Chain}

\begin{figure}
\centerline{
    \includegraphics[width=\figurewidth]{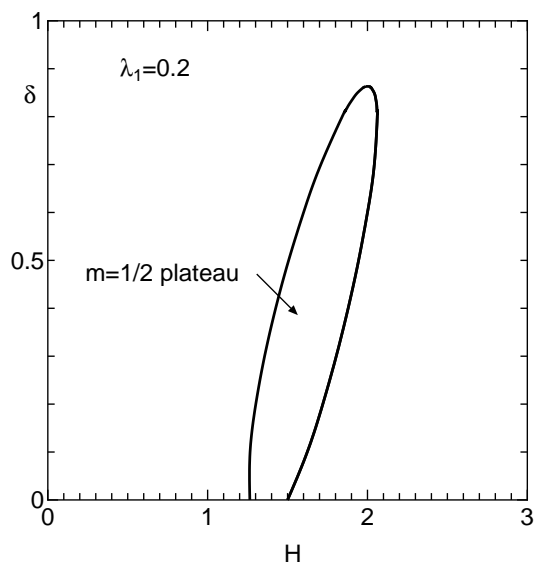}
}
  \caption{Magnetic phase diagram of the $S=1/2$ single dimerized zigzag Heisenberg chain on the $H-\delta$ plane for $\lambda_1=0.2$.}
  \label{1dfld}
\end{figure}

Figure \ref{1dfld} shows the magnetic phase diagram of the $S=1/2$ one-dimensional dimerized zigzag chain ($\lambda_2=0$) on the $H-\delta$ plane determined by the present method for $\lambda_1=0.2$. Compared to the numerical results (Fig. 3 of ref. \citen{tone1}), the BOMFA overestimates the width of the plateau. Especially, the plateau width remains finite even for $\delta=0$ where no plateau is observed in numerical calculation. Nevertheless, the approximate position of the plateau and the range of $\delta$ for which the plateau appears is fairly well reproduced.

\begin{figure}
\centerline{
    \includegraphics[width=\figurewidth]{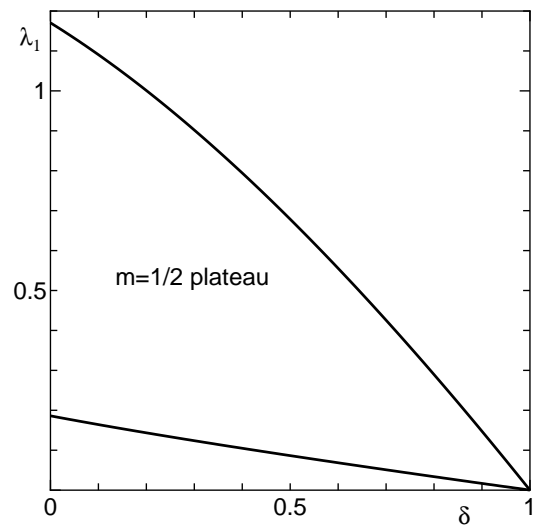}
}
  \caption{Ground state phase diagram of the one-dimensional dimerized zigzag chain for $m=1/2$ on the $\delta - \lambda_1$ plane. The magnetic field $H$ is scaled by $J$.}
\label{zig1dpha}
\end{figure}

The latter feature is visible more clearly in Fig. \ref{zig1dpha} which shows the ground state phase diagram  on the $\delta-\lambda_1$ plane. The isolated dimer limit, which is the starting point of the BOMFA, is characterized by $\lambda_1=0$ and $\delta=1.0$. For finite $\lambda_1$ and/or $1-\delta$, the interdimer interaction is switched on and the many body effect stabilizes the $m=1/2$ plateau state accompanied by STSB. Comparing the present results with the numerical phase diagram by Tonegawa and coworkers, \cite{tone, tone1}(Fig. 1 of ref. \citen{tone1}) the plateau region is well reproduced by the BOMFA. This implies that the present approximation takes into account the many body effect appropriately.

\subsection{Two Dimensional Coupled Dimerized Zigzag Chain}

The results of the preceding subsection indicates that the BOMFA gives reliable results even for the one-dimensional case in which the mean field type approximation is expected to be rather poor.  Encouraged by this success, we further investigate the effect of the interchain coupling in this subsection. 

\begin{figure}
\centerline{
    \includegraphics[width=\figurewidth]{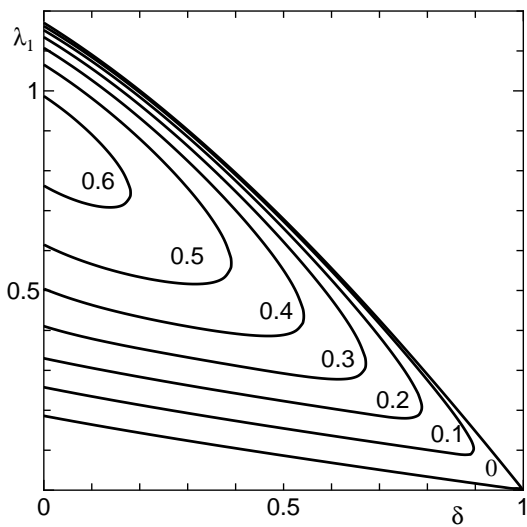}
}
  \caption{The ground state phase diagram of the two dimensionally coupled $S=1/2$ dimerized zigzag Heisenberg chains for $m=1/2$. The region encircled by each curve and $\lambda_1$-axis is the plateau region. The values of $\lambda_{2}$ are indicated for each line.}
\label{2dzig}
\end{figure}
The $m=1/2$ ground state phase diagram on the $\delta-\lambda_{1}$ plane is shown for various values of $\lambda_{2}$ in Fig. \ref{2dzig}. It is seen that the interchain interaction suppresses the $m=1/2$ plateau. Actually, no plateau appears for $\lambda_2 > 0.634$. This conclusion is physically reasonable because the present plateau is understood as the solidification of the singlet and triplet pairs which alternate on the strong bonds.\cite{totsuka} The interchain coupling tends to align the spins and destroy the singlet pair formation. Therefore the present plateau is destabilized by the interchain interaction.  Considering that the mean field approximation becomes more reliable as the dimensionality becomes higher, we expect the present calculation is reliable in the coupled chain system.

\begin{figure}
\centerline{
    \includegraphics[width=\figurewidth]{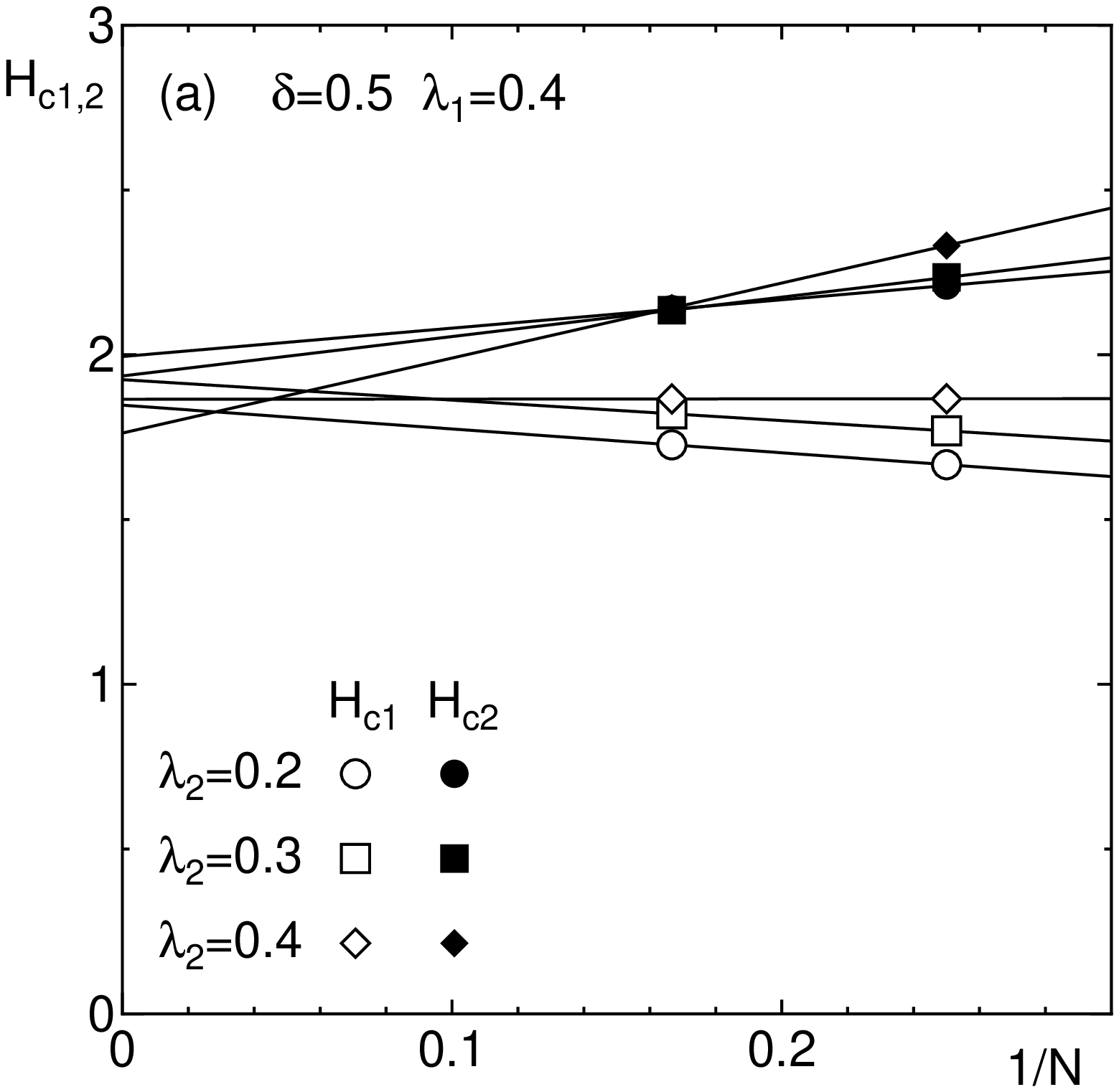}}
\centerline{
    \includegraphics[width=\figurewidth]{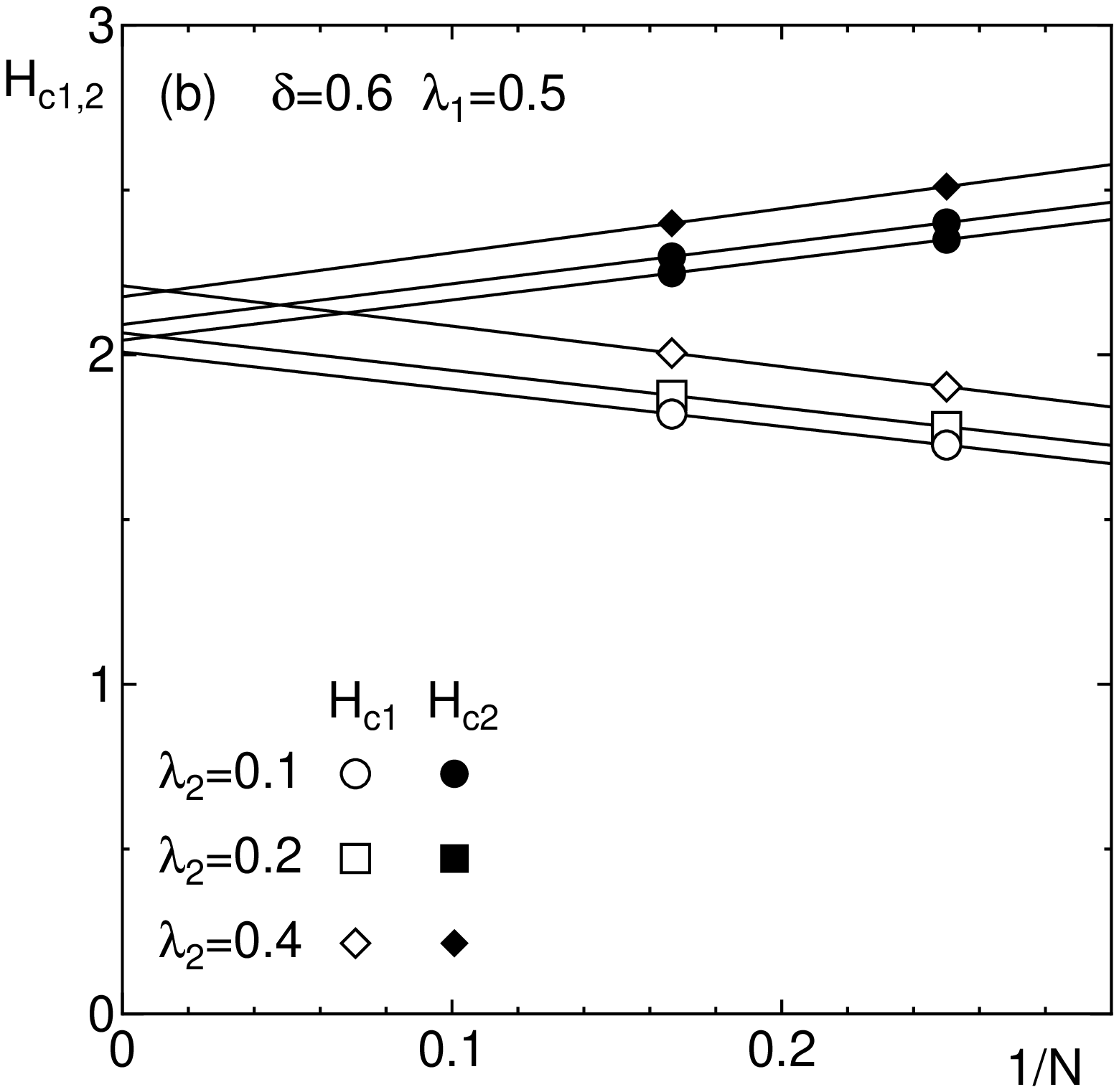}
}
  \caption{The magnetic fields at lower end  $H_{c1}$ and upper end $H_{c2}$ of the $m=1/2$ plateau of the $S=1/2$ coupled dimerized zigzag Heisenberg chains for the clusters with $N=\Nc=4$ and 6 plotted against $1/N$ for (a) $(\delta, \lambda_1)=(0.5, 0.4)$ and (b) (0.6, 0.5). The solid line is the linear extrapolation in $1/N$.  The magnetic field is scaled by $J$.}
\label{nume}
\end{figure}

To confirm the above conclusion numerically, we have carried out the numerical diagonalization calculation for $\Nc=N=4$ and 6 clusters.  Figure \ref{nume} shows the $N(=\Nc)$-dependence of the magnetic fields at the lower end $H_{c1}$ and upper end $H_{c2}$ of the $m=1/2$ plateau for (a) $(\delta, \lambda_1)=(0.5, 0.4)$ and (b) $(\delta, \lambda_1)=(0.6, 0.5)$, respectively.  Of course, the system size dependence of  $H_{c1}$ and $H_{c2}$ is not clear, so that there is large ambiguity in the extrapolation to $N \rightarrow \infty$ from only two values of $N$. Therefore the absolute values of $H_{c1}$ and $H_{c2}$ are not reliable. However, it is clear from  Fig. \ref{nume} that the extrapolated values of the plateau width decrease with $\lambda_2$. It is also checked that this tendency is insensitive to the way of extrapolation. Therefore the numerical diagonalization results support the validity of the BOMFA result.

\section{Summary and Discussion}

In the present work, we have investigated the $m=1/2$ plateau of the coupled dimerized zigzag chain by applying the BOMFA to the plateau state. It is found that the $m=1/2$ plateau associated with STSB can be described by BOMFA. In the single dimerized zigzag chain, the ground state phase diagram obtained by the numerical calculation is reproduced quantitatively well. In the coupled dimerized zigzag chains, the plateau is found to be suppressed by the interchain coupling. This is also consistent with the numerical diagonalization results.

Similar conclusion is also obtained by Kolezhuk\cite{kole} several years ago. However, his calculation treats the interdimer interaction only perturbatively and no quantitative phase diagram is obtained. Furthermore, the effective kinetic energy of triplet pairs are assumed to be much smaller than the interaction energy between these pairs, although both energies are in fact the same order in $\lambda_1, \lambda_2$ and $1-\delta$. In the present treatment, the both energies are equally taken into account. 

Sommer and coworkers\cite{svb} and Matsumoto and coworkers\cite{mnrs, matsu} also applied BOMFA to the magnetization process of the quantum magnets. However, these authors concentrated on the non-plateau part of the magnetization curve and explained the field induced transverse antiferromagnetic ordering in two and three dimensional models. The simple-minded application of their method to the one-dimensional case also leads to the transverse ordering. This is obviously unphysical, because the ground state of the quantum spin chains in the non-plateau region should be the Luttinger liquid state. Furthermore, the plateau associated with the STSB cannot be described by their scheme, because STSB occurs only on the magnetization plateau. On the other hand, our approach concentrate on the plateau state and is not suitable for the description of the non-plateau part of the magnetization curve. It is, however, applicable to the one-dimensional case and can describe the plateau state with STSB. Therefore we may conclude that our method and the method of refs. \citen{matsu, svb, mnrs} are complementary to each other.

For $\delta=0$ and $\lambda_2=0$, Okunishi and Tonegawa\cite{oku, oku2} have shown that the $m=1/3$ plateau appears in the strongly frustrated regime. For $\delta=0, \lambda_1=\lambda_2=1$, the present model reduces to the triangular lattice for which the $m=1/3$ plateau is believed to be present.\cite{nm, sm} In this context, it is interesting to investigate the interplay of two dimensionality, dimerization and frustration in the $m=1/3$ plateau problem. We have also tried to apply the present method to the $m=1/3$ plateau. However, the width of the plateau region on the $\delta-\lambda_1$ plane determined by the BOMFA is substantially wider than the  numerical results by Okunishi and Tonegawa\cite{oku, oku2} for $\delta=0$  and preliminary DMRG calculation by one of the authors(KH) for $\delta \ne 0$. The origin of this discrepancy is the following. As discussed by Okunishi and Tonegawa, \cite{oku, oku2} this $m=1/3$ plateau is basically of classical origin with up-up-down local spin configuration realized in the Ising limit rather than the quantum mechanical plateau resulting from the singlet pair formation on the strong bonds. Therefore the present approach based on the singlet pair formation is not suitable for the description of the $m=1/3$ plateau. The same is true for the triangular  lattice ($\delta=0, \lambda_1=\lambda_2=1$)  which also has the $m=1/3$ plateau.\cite{nm, sm} Therefore in order to apply the present method to the $m=1/3$ plateau, it would be necessary to refine the approximation so as to incorporate the effect of the ground state polarization of the singlet sites. This is left for future studies.

In conclusion, the BOMFA is a promising tool for the investigation of the magnetization plateau of quantum origin including those induced by many body effect associated with the translational symmetry breakdown. The application of the present method to wide range of models is hoped in the future.

\noindent
{\bf Ackowledgement}

The authors thank T. Yamamoto for enlightening comments to the earlier version of the present manuscript. The numerical diagonalization calculation in this work has been carried out using the facilities of the Supercomputer Center, Institute for Solid State Physics, University of Tokyo and the Information Processing Center, Saitama University.  This work is supported by a Grant-in-Aid for Scientific Research from the Ministry of Education, Culture, Sports, Science and Technology, Japan.

\end{document}